# Multilevel AlScN Ferroelectric Diode Crossbar Arrays with Sub-200 nm Cells for Selector-Free Nonvolatile Memory

Kwan-Ho Kim,[1†] Zekun Hu,[1†] Krishna Karthikeya Chemudupati,[1,2†] Taarana Jammula,[1] Yinuo Zhang,[1] Rajeev Kumar Rai,[3] Dhiren K. Pradhan,[1] Zirun Han,[1] Eric A. Stach,[3] Roy H. Olsson III,[1] and Deep Jariwala[1,*]

[1]*Department of Electrical and Systems Engineering, University of Pennsylvania, Philadelphia, Pennsylvania 19104, United States*

[2]*Department of Physics and Astronomy, University of Pennsylvania, Philadelphia, Pennsylvania 19104, USA*

[3]*Department of Materials Science and Engineering, University of Pennsylvania, Philadelphia, Pennsylvania 19104, USA*

[†]*These authors equally contributed to this work.*

[*]Author to whom correspondence should be addressed: dmj@seas.upenn.edu

**ABSTRACT**

Passive crossbar arrays are a leading route to high-density nonvolatile memory and in-memory computing, but without a selector the memory element itself must supply the nonlinearity that suppresses sneak-path currents. Here we scale aluminum scandium nitride (AlScN) ferroelectric diodes (FeDs) in both dimensions that matter - the ferroelectric is thinned to 10 nm and the junction diameter is reduced to 100 nm. We further integrate 200 nm cells into a 2-kilobyte (128 × 128) passive crossbar. At a junction area of 0.0314 $\mu m^2$ this is a ~625× reduction in cell area, at equal array capacity, relative to kilobyte-scale AlScN FeD crossbars built on 5 μm cells. The 10 nm $Al_{0.64}Sc_{0.36}N$ layer reverses its polarization under pulses of 50 ns full width at half maximum at a pulse-energy density of ~16 pJ $\mu m^{-2}$, corresponding to ~0.5 pJ per 200 nm cell. Pulsed programming at 17.5 V and 1 kHz yields an ON/OFF ratio of $4.2 \times 10^3$ in a 200 nm cell, while DC characterization of 283 randomly addressed cells of the 2-kB array gives ON/OFF and rectification ratios at 5 V of 43 ± 33 and 61 ± 74, from which we derive the sub-array size that selector-free reading currently supports. Eight conductance levels (three bits) written into a 200 nm array cell remain resolvable for $10^3$ s, and the two extreme states are stable beyond $5 \times 10^4$ s. Mapping the measured, nonuniformly spaced levels onto a ResNet-20 network recovers 91.7 ± 0.1 % CIFAR-10 accuracy after quantization-aware training, against a 92.2 ± 0.1 % full-precision baseline. AlScN FeDs thus combine nanosecond switching, three-bit storage and sub-0.05 $\mu m^2$ cells in a BEOL-compatible, selector-free crossbar.

## INTRODUCTION

The computational demands of modern artificial intelligence have grown at a rate that increasingly challenges conventional computing hardware. Training compute for machine-learning models has doubled approximately every six months since about 2010, roughly four times faster than in the preceding six decades and far faster than transistor density has grown under Moore's law[1]. Furthermore, conventional computing systems, built on the von Neumann model, require data to be continuously shuttled between physically separated memory and processing units. For neural network inference, where the dominant operation is the multiply-accumulate, a substantial fraction of the energy is associated with data movement rather than arithmetic operations. As a result, transistor shrinkage alone cannot close a gap that originates from physical separation rather than device dimensions. The rapid scaling of large language models and other AI architectures underscores the urgent need for dense, energy-efficient non-volatile memory compatible with CMOS back-end-of-line (BEOL) processing[2].

BEOL-compatible in-memory computing has recently attracted considerable attention as a strategy to overcome the von Neumann bottleneck by performing computation directly within memory arrays[3]. Crossbar architectures based on two-terminal non-volatile memory devices are particularly attractive owing to their compact footprint, scalability, and ability to execute analog matrix-vector operations in situ[4]. Recent demonstrations using resistive and phase-change memory technologies have highlighted the potential of such architectures for highly energy-efficient computing[3, 5]. However, realizing large-scale crossbar arrays requires memory devices that combine a large ON/OFF ratio for reliable state discrimination, multilevel programmability for increased storage density and analog weight representation, and intrinsic current rectification to mitigate parasitic leakage pathways. In passive crossbar architectures, sneak-path currents can severely degrade read accuracy and limit array scalability[6-8]. The requirement this places on the memory element is quantitative: for the worst case of reading a high-resistance cell embedded in an all-low-resistance background, the read margin of an $N \times N$ passive array is approximately $(1 - 1/r)/[1 + (N - 1)/NL]$, where $r$ is the ON/OFF ratio and NL is the current nonlinearity at the bias seen by the half-selected cells. Retaining a conventional 10 % margin at $N = 128$ therefore requires $NL \gtrsim 15$, and at $N = 512$ requires $NL \gtrsim 60$ - specifications that no linear resistive element can meet. Although additional selector devices can suppress these unwanted currents, they increase fabrication complexity, cell footprint, and integration cost. Consequently, self-rectifying memory devices capable of selector-free operation have emerged as an attractive pathway toward dense and scalable crossbar arrays (Fig. 1a).

Ferroelectric diodes (FeD) offer an attractive route toward selector-free crossbar memories by combining nonvolatile resistance switching and intrinsic rectification within a single two-terminal device[9]. The polarization-controlled modulation of interfacial barriers suppresses sneak-path currents without requiring additional selector elements, while supporting multilevel conductance states for high-density memory storage. However, maintaining large rectification ratios, reliable multilevel programmability, and device-to-device uniformity across large-scale passive crossbar arrays remains a significant challenge.

Among emerging ferroelectric materials, AlScN is particularly attractive for scalable FeD architectures owing to its large remanent polarization, thermal robustness, and compatibility with back-end-of-line processing.[10, 11] Unlike hafnium oxide ferroelectrics, whose performance relies on a metastable ferroelectric phase, ferroelectricity in AlScN originates from the stable wurtzite crystal structure, enabling reliable operation across a broad processing and temperature window. These attributes make AlScN a promising platform for achieving the large polarization-driven barrier modulation required for high ON/OFF ratios, strong self-rectification, and selector-free operation in dense crossbar arrays.[12] Progress on AlScN FeDs has, however, advanced along two tracks that have not yet met. Aggressive lateral scaling has been demonstrated only on standalone devices - most recently to 40 nm diameter, with an ON/OFF ratio above 60 and three-bit retention beyond $4 \times 10^4$ s at 85 °C - whereas the kilobyte-scale, selector-free AlScN FeD crossbars reported to date are built on 5 µm cells and reach a memory density of only ~$2.5 \times 10^3$ bit mm$^{-2}$. Bridging the two is not merely a lithography exercise. Shrinking the junction reduces the read current in proportion to area, so that a 200 nm cell delivers ~$10^{-10}$ A in the low-resistance state and pushes the high-resistance state toward the noise floor of a conventional parameter analyzer; simultaneously, the depolarizing field associated with a 4 nm dielectric interlayer on a 10 nm ferroelectric grows as the ferroelectric is thinned, and the sidewall of a nanoscale via becomes a competitive leakage path. Whether large ON/OFF ratios, self-rectification and multilevel programmability survive simultaneous vertical and lateral scaling inside a working passive array is therefore an open question, and it is the question we address here.

Here we report a 2-kilobyte (128 × 128) selector-free AlScN FeD crossbar array built on 200 nm diameter cells, together with standalone and 8-byte-array devices scaled to 100 nm. At a junction area of 0.0314 $\mu m^2$, corresponding to a projected $4F^2$ density of 6.3 Mbit $mm^{-2}$ at F = 200 nm and 25 Mbit $mm^{-2}$ at F = 100 nm. The 10 nm $Al_{0.64}Sc_{0.36}N$ layer switches under 50 ns pulses at ~16 pJ $\mu m^{-2}$, i.e. ~0.5 pJ for a 200 nm cell and ~0.13 pJ for a 100 nm cell. Pulsed programming at 17.5 V and 1 kHz yields an ON/OFF ratio of $4.2 \times 10^3$ in a 200 nm cell, while DC characterization of 283 randomly addressed cells of the 2-kB array gives an ON/OFF ratio of 43 ± 33 and a rectification ratio of 61 ± 74 at 5 V. Eight conductance levels programmed into a 200 nm array cell remain distinguishable for $10^3$ s, and the two states are stable beyond $5 \times 10^4$ s. Finally, we quantify what the measured, nonuniformly spaced levels cost a ResNet-20 network on CIFAR-10 and show that quantization-aware training recovers accuracy to within 0.5 % of the full-precision baseline. Together these results define both the reach and the present limits of selector-free AlScN FeD crossbars at sub-200 nm cell dimensions.

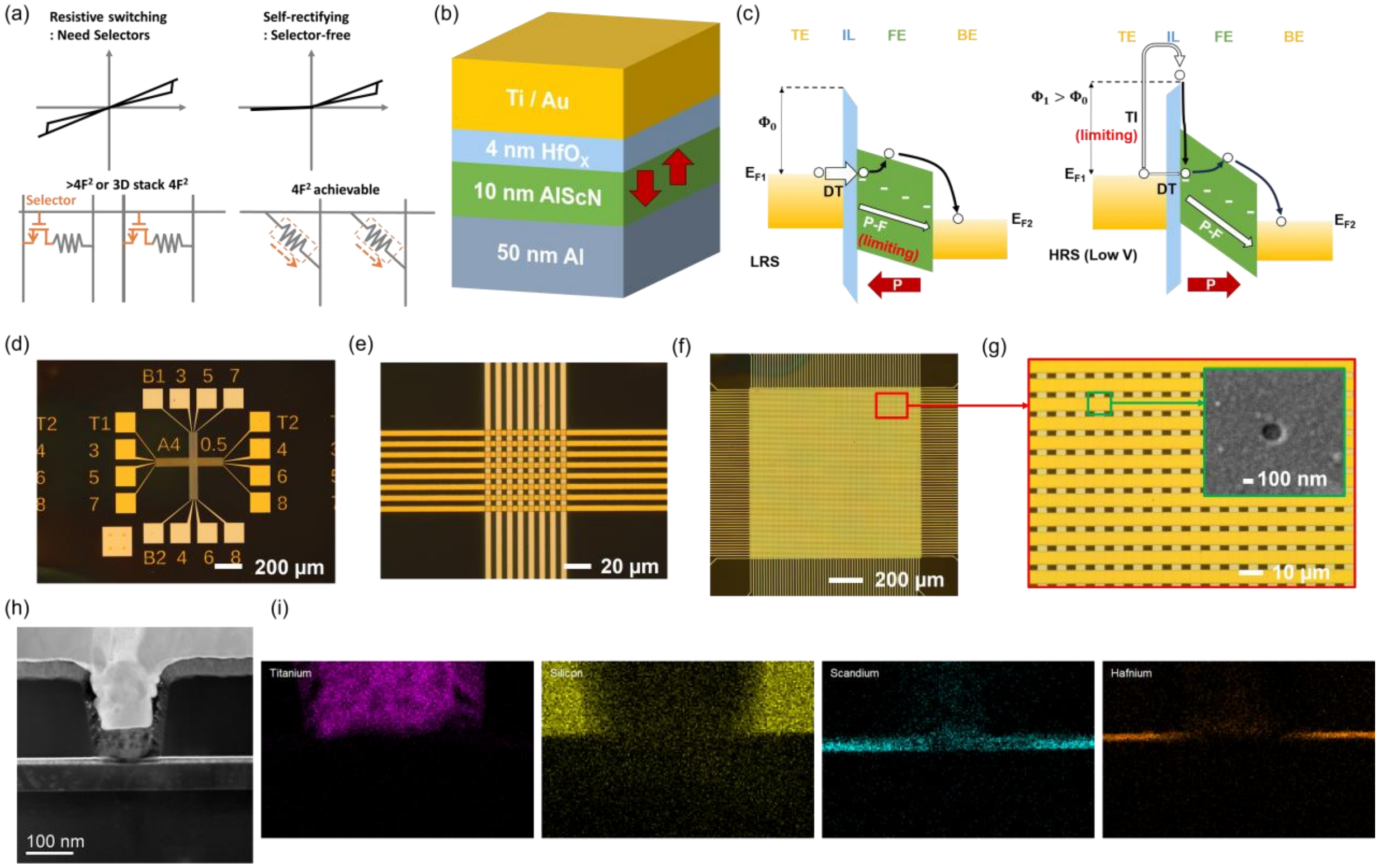

**Figure 1. Device concept, crossbar architecture, and structural characterization of AlScN ferroelectric diodes.** (a) Comparison between conventional resistive-switching crossbars requiring additional selectors and self-rectifying ferroelectric-diode crossbars enabling selector-free operation with a projected $4F^2$ cell footprint. (b) Schematic of the metal-ferroelectric-insulator-metal device stack consisting of a 50 nm Al bottom electrode, 10 nm AlScN ferroelectric layer, 4 nm $HfO_x$ interlayer, and Ti/Au top electrode. (c) Schematic energy-band diagrams illustrating polarization-controlled carrier transport in the LRS and HRS under a low read bias. Polarization reversal changes the effective injection barrier and the relative contributions of DT and P-F transport. (d-f) Optical micrographs of a standalone device, an 8-byte crossbar array, and a 2-KB crossbar array, respectively. (g) Magnified view of the nanoscale crossbar structure; the inset shows an individual junction of the 2-kB array with a nominal diameter of 200 nm (scale bar, 100 nm). (h) Cross-sectional transmission electron micrograph of the via-defined FeD junction from an 8-B crossbar array. (i) Corresponding elemental maps of Ti, Si, Sc, and Hf, confirming preservation of the vertically integrated device structure after fabrication.

## RESULTS AND DISCUSSION

### Scalable AlScN Ferroelectric Diode Crossbar Architecture

The FeD devices in this work consist of a 50 nm thick Al bottom electrode serving as the word line, over which a 10 nm thick $Al_{1-X}Sc_XN$ (X = 0.28, 0.32, and 0.36) ferroelectric layer is sputter-deposited. A 4 nm $HfO_x$ interlayer (IL) is subsequently deposited via atomic layer deposition (ALD) on top of the AlScN. The schematic of this MFIM stack is shown in Figure 1b. The adoption of $HfO_X$ as the IL, rather than $AlO_X$, is motivated by its higher dielectric constant, which reduces the depolarization field in the scaled 10 nm AlScN layer and thereby improves retention.[12] The top electrodes, which also serve as the bit lines of the crossbar array, were defined by electron-beam lithography and metallized with Ti/Au, completing the metal–ferroelectric–insulator–metal (MFIM) device structure. Both the word/bit lines and pitch were designed with a width of 8 μm to ensure reliable fabrication and high array yield. This relatively large line width was selected for fabrication robustness and does not represent an intrinsic scaling limitation of the FeD device architecture, as further supported by the nanoscale device scaling results presented in this work.

To evaluate the scalability of AlScN FeD memories from individual devices to large-scale integration, we progressively expanded the architecture from standalone FeDs to 8-byte (8 × 8, 64-cell) and ultimately 2-kilobyte (128 × 128, 16,384-cell) selector-free crossbar arrays. The intrinsic self-rectifying behavior of the MFIM FeD eliminates the need for dedicated selector devices, enabling a true $4F^2$ cell footprint and high-density passive crossbar integration (Fig. 1a). The operating principle of the AlScN FeD is illustrated in Fig. 1c, where polarization reversal modulates the electrostatic potential across the $HfO_X$ interfacial layer and AlScN ferroelectric layer. In the low-resistance state, polarization lowers the effective barrier for carrier injection and enables bulk-limited Poole-Frenkel transport through AlScN, whereas in the high-resistance state the increased barrier suppresses low-bias current and limits injection through direct tunneling or thermionic emission (Supporting Information Note S1 and Figure S1). This polarization-dependent control of carrier injection and transport gives rise to both nonvolatile resistance switching and intrinsic rectification, enabling selector-free memory operation in the crossbar array.

Optical micrographs of the fabricated structures are shown in Fig. 1d-g. Figure 1d presents the standalone FeD test structure used for device-level characterization. The crossbar architecture was subsequently scaled to an 8-B array and further expanded to a 2-KB array, as shown in Fig. 1e and f. Figure 1g highlights the highly scaled crossbar implementation, with the inset SEM image confirming successful fabrication of the 200 nm junctions used in the 2-kB array; junctions down to 100 nm in diameter were fabricated and characterized in the 8-B arrays (Fig. 3g). Cross-sectional structural analysis further confirms the successful fabrication of the scaled MFIM junctions. The cross-sectional TEM image in Fig. 1h shows a well-defined via-contact geometry with the top electrode contacting the underlying FeD stack through the patterned nanoscale opening. Elemental mapping in Fig. 1i verifies the intended vertical material distribution, including the Ti top electrode, Sc-containing AlScN ferroelectric layer, and Hf-containing interfacial layer. The spatially confined elemental profiles confirm that the active junction is defined by the via geometry and that the ferroelectric/interfacial-layer stack is preserved after device fabrication.

**Ferroelectric Switching Characteristics of 10 nm $Al_{0.64}Sc_{0.36}N$**

We first establish the intrinsic ferroelectric switching characteristics of the 10 nm $Al_{0.64}Sc_{0.36}N$ layer, which provides the polarization modulation required for subsequent FeD operation. Figure 2a shows the current-density-electric-field (J-E) characteristics of a representative device with a radius of 5 μm, measured at 10 kHz. Hysteresis between the increasing- and decreasing-field branches is resolved in both polarities, with the branch separation centred near ±7 MV $cm^{-1}$, which we take as the coercive field of this 10 nm layer. We note that at this thickness the switching current is not resolved as a discrete peak above the leakage background, so Fig. 2a is used only to locate $E_c$; the switchable charge itself is quantified by the PUND measurements below.

The switching polarization is further evaluated using positive-up-negative-down (PUND) measurements. Figure 2b shows the time-domain current response under the P/U/N/D pulse sequence, where pronounced switching-current peaks are observed during the P and N pulses and are substantially suppressed during the corresponding U and D pulses. The resulting P-U and N-D charge differences increase systematically with programming voltage, as shown in Fig. 2c, confirming a large electrically switchable charge response in the ultrathin AlScN film. The charge difference reaches ~800 μC $cm^{-2}$ at 9.3 V, three- to fourfold larger than the $2P_r$ of 200-270 μC $cm^{-2}$ expected for $Al_{0.64}Sc_{0.36}N$, and it does not saturate over the accessible voltage range. The excess is leakage that the U and D pulses do not cancel, and the quantity plotted must therefore be read as a total switching-related charge, not as $2P_r$. This interpretation is corroborated by Fig. 2g, in which the U and D transients are almost as large as the P and N transients.

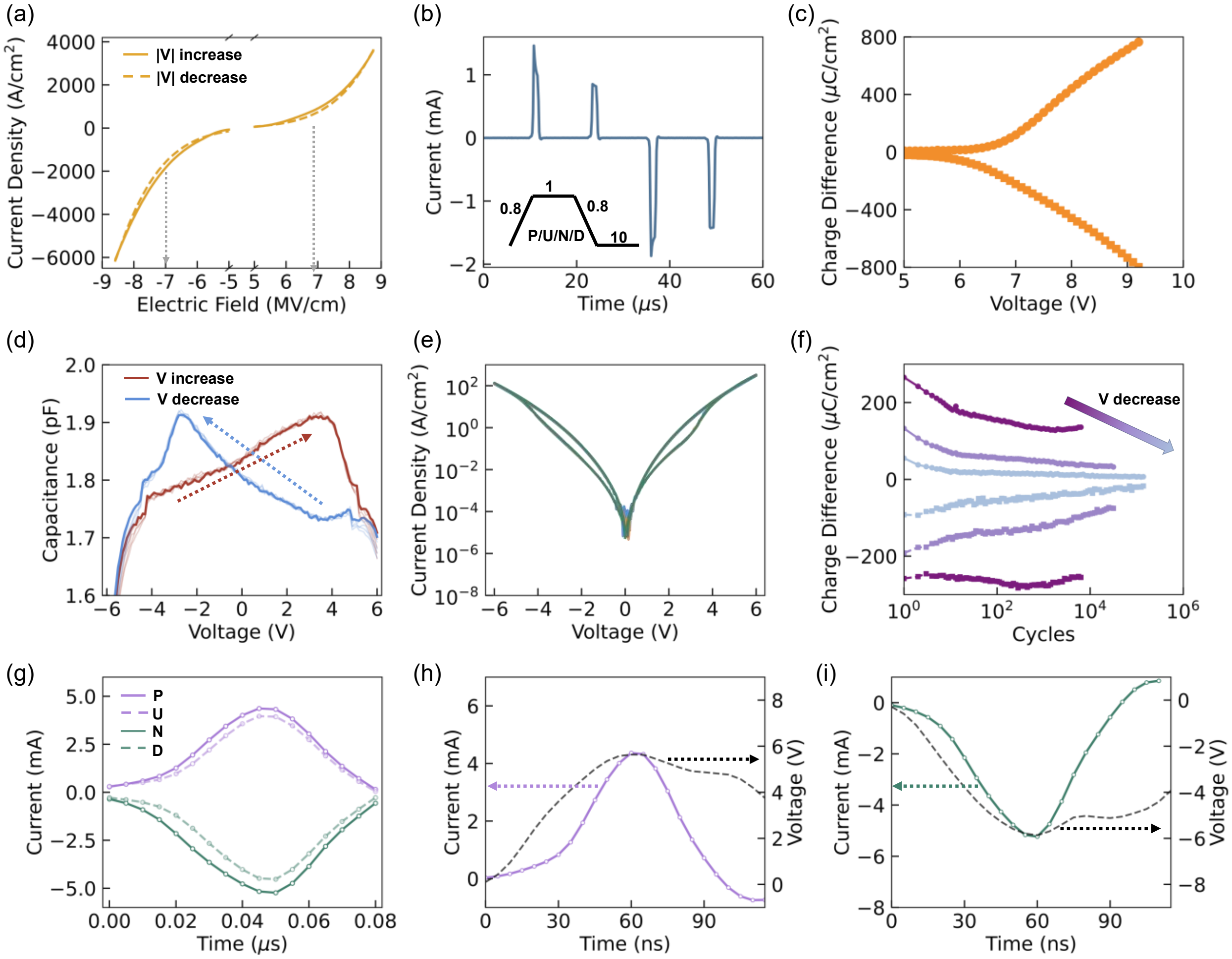

**Figure 2. Ferroelectric switching, endurance, and nanosecond polarization dynamics of 10 nm $Al_{0.64}Sc_{0.36}N$.**

**(a)** Current density-electric field (J-E) characteristics of a 10 nm $Al_{0.64}Sc_{0.36}N$ ferroelectric capacitor showing bipolar polarization switching under increasing and decreasing electric-field sweeps. **(b)** Representative time-domain PUND current response with the applied P/U/N/D pulse sequence shown schematically. Pronounced current peaks during the P and N pulses relative to the U and D pulses identify polarization-switching contributions. **(c)** PUND-extracted P-U and N-D charge differences as a function of programming voltage. **(d)** Capacitance-voltage hysteresis measured under opposite voltage-sweep directions, showing polarization-dependent dielectric modulation. **(e)** Repeated current density-voltage characteristics demonstrating reproducible area-distributed electrical transport. **(f)** Endurance characteristics extracted from PUND charge differences at different programming voltages, showing persistent separation of the positive and negative switching branches during cycling. **(g)** Positive and negative nanosecond switching-current transients. **(h,i)** Time-resolved positive and negative switching responses, respectively, together with the applied voltage waveform. Polarization reversal is achieved using voltage pulses with FWHM as short as 50 ns, corresponding to an estimated switching-energy density of approximately 16 pJ $\mu m^{-2}$.

**Scalability of Self-Rectifying AlScN Ferroelectric Diodes**

Independent capacitance and transport measurements further support the ferroelectric origin of the electrical response. The capacitance-voltage characteristics in Fig. 2d exhibit pronounced hysteresis and voltage-shifted capacitance maxima between opposite sweep directions, consistent with polarization reversal and the associated field-dependent dielectric response. Meanwhile, the repeated current-density-voltage characteristics in Fig. 2e show reproducible area-distributed transport without signatures characteristic of localized filamentary switching. Together, the J-E, PUND, C-V, and J-V measurements establish robust polarization-controlled switching in the 10 nm $Al_{0.64}Sc_{0.36}N$ layer.

The switching characteristics remain stable under repeated electrical stress. Figure 2f summarizes the PUND charge difference measured during endurance cycling at multiple programming amplitudes. Positive and negative branches remain separated over the accessible cycling range at the higher amplitudes, whereas at the lowest amplitude they converge to within the measurement scatter after ~$10^3$ cycles; larger programming voltages produce correspondingly larger switching-charge windows but terminate earlier in hard breakdown. We emphasize that Fig. 2f reports the fatigue of the ferroelectric capacitor and not the write endurance of the FeD memory window, which is discussed separately below.

Importantly, the polarization reversal also persists under nanosecond excitation. The short-pulse measurements in Figs. 2g-i resolve both positive and negative switching-current transients under rapidly varying voltage pulses. Reliable polarization reversal is obtained using voltage pulses with a full width at half maximum (FWHM) as short as 50 ns, while atomistic calculations describing wurtzite polarization reversal as a column-by-column process propagating from a slow, fractal-like domain wall suggest that the intrinsic switching dynamics may extend into the sub-nanosecond regime. Integrating V(t)I(t) over the pulse gives a total pulse-energy density of approximately 16 pJ $\mu m^{-2}$. Because the non-switching (U/D) contribution has not been subtracted, this is an upper bound on the switching energy; scaled to the array cells it corresponds to ~0.50 pJ per 200 nm cell and ~0.13 pJ per 100 nm cell. This value is consistent with the thermodynamic estimate $2P_rE_ct \approx 10$ pJ $\mu m^{-2}$ for $2P_r = 200$ $\mu C$ $cm^{-2}$, $E_c = 5$ MV $cm^{-1}$ and $t = 10$ nm, and it is intrinsically one to two orders of magnitude larger than for $HfO_2$-based ferroelectrics because of the much larger coercive field of the wurtzite system. These results demonstrate that the scaled AlScN ferroelectric combines robust polarization, cycling stability, and nanosecond switching with low switching energy, providing a strong device-level basis for high-speed nonvolatile FeDiode memory operation.

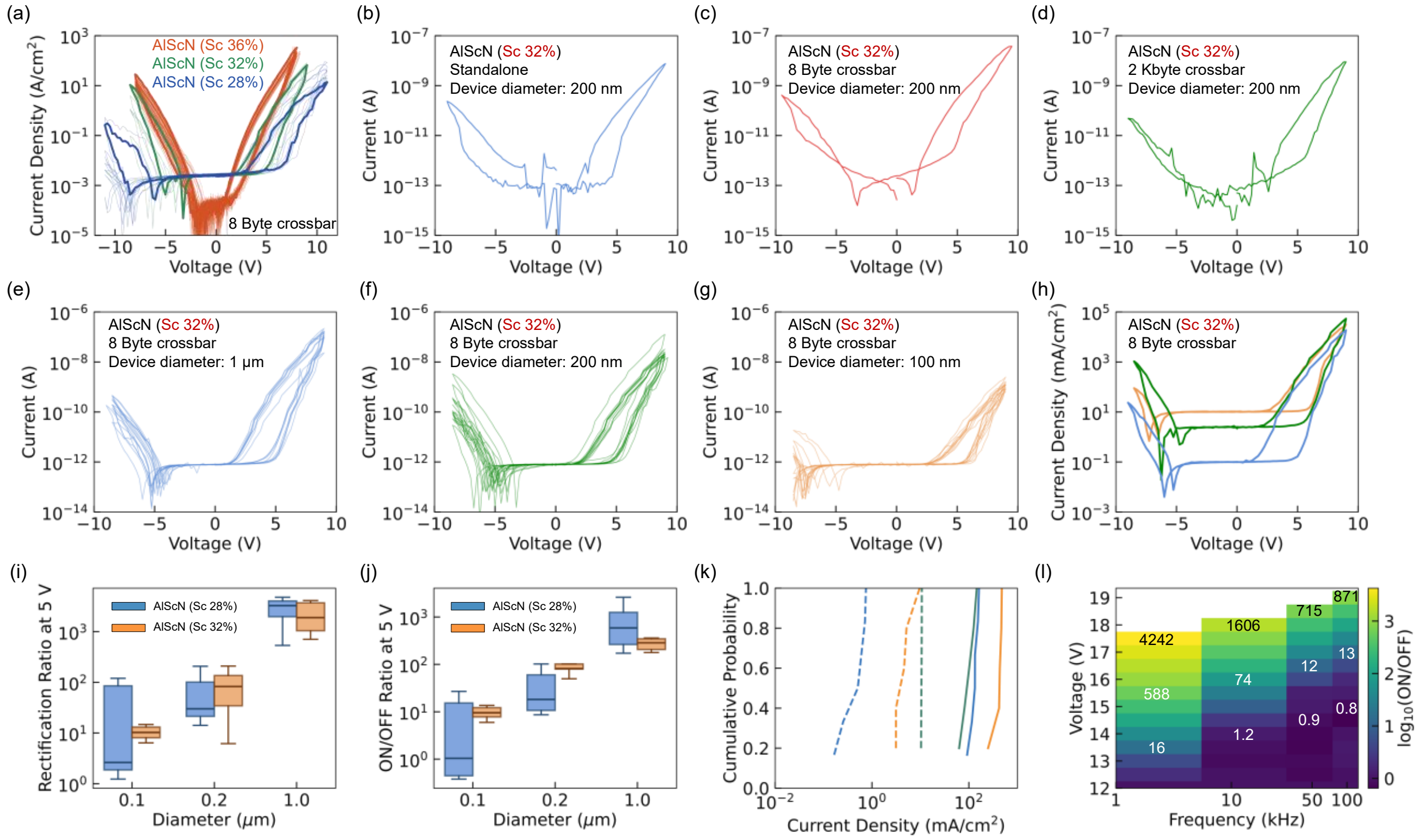

**Figure 3. Electrical scalability and pulse-programming characteristics of self-rectifying AlScN ferroelectric diodes.** (a) Current density-voltage characteristics of devices containing 28%, 32%, and 36% Sc, showing the influence of Sc composition on switching voltage, leakage current, and memory window. (b-d) Representative current-voltage characteristics of a 200 nm diameter standalone device, a 200 nm diameter device integrated in an 8-B crossbar, and a 200 nm device integrated in a 2-KB crossbar, respectively. (e-g) Repeated current-voltage characteristics of devices with diameters of 1 μm, 200 nm, and 100 nm, respectively, measured within the 8-B crossbar architecture. (h) Area-normalized current-density characteristics for the different device dimensions, showing comparable ON-state current densities and an OFF-state current approaching the measurement floor in the smallest devices. (i,j) Statistical distributions of the rectification ratio and ON/OFF ratio, respectively, measured at 5 V for 28% and 32% Sc devices with diameters ranging from 100 nm to 1 μm. (k) Cumulative probability distributions of the ON- and OFF-state current densities for the different device dimensions. (l) Heat map of $\log_{10}$(ON/OFF) as a function of programming voltage and pulse frequency, demonstrating pulse-programmable switching up to 100 kHz.

Fig. 3 evaluates the electrical scalability of AlScN FeDs across Sc composition, array size, and device dimension. We first fabricated 8-B crossbar arrays using AlScN films with Sc contents of 28%, 32%, and 36% to identify an optimized composition for scaled array operation (Fig. 3a). As the Sc content increases, the reduced coercive field lowers the operating voltage range; however, the 36 % Sc devices exhibit a reduced memory window, which we attribute to the higher leakage of the lower-bandgap. The 32% Sc composition provides the best balance between low-voltage operation, large ON/OFF ratio, and reliable self-rectifying switching, and was therefore used for the subsequent standalone, 8-B, and 2-KB array demonstrations[12]. Figures 3b-d compare representative I-V characteristics measured from a standalone device, an 8-B crossbar array, and a 2-KB crossbar array, respectively. In all cases, the devices exhibit hysteretic and asymmetric switching behavior characteristic of polarization-controlled transport in self-rectifying FeDs. The DC ON/OFF ratio measured at the array level falls from 275 (8-B, 28 % Sc) and 156 (8-B, 32 % Sc) to 48 in the 2-kB array, and the rectification ratio from 1481 (8-B, 32 % Sc) to an average of 232, which we attribute to the sneak-path contribution of the 16,383 unselected cells that share the selected word and bit lines. The memory window and the rectifying asymmetry are nonetheless clearly preserved in the 2-kB array. These results demonstrate that intrinsic selector-free operation is maintained across progressively larger AlScN FeD crossbar arrays.

To further investigate device scaling, Figures 3e-g present I-V characteristics measured from multiple devices with diameters of 1 μm, 200 nm, and 100 nm, respectively. Across all dimensions, the devices exhibit reproducible ferroelectric switching with well-defined high- and low-resistance states. Self-rectifying behavior and nonvolatile resistance modulation persist down to the smallest device diameter of 100 nm, indicating that polarization-controlled barrier modulation survives aggressive lateral scaling - but not without cost. For the 32 % Sc composition the median ON/OFF ratio at 5 V falls from $\sim 3 \times 10^2$ at 1 μm to $\sim 9 \times 10^1$ at 200 nm and $\sim 1 \times 10^1$ at 100 nm, and the median rectification ratio from $\sim 2 \times 10^3$ to $\sim 4 \times 10^1$ and $\sim 1 \times 10^1$ over the same range (Fig. 3i,j). These observations establish that the AlScN FeD architecture remains functional across both increasing array size and decreasing device dimensions, while quantifying the price paid in memory window and rectification. In an ideal crossbar layout with a $4F^2$ cell footprint, the 200 nm cells used in the 2-kB array correspond to a projected density of 6.3 Mbit mm$^{-2}$, and the 100 nm cells demonstrated in the 8-B arrays to 25 Mbit mm$^{-2}$.

As the device diameter is reduced from 1 μm to 100 nm, increased device-to-device variation is observed in both the high- and low-resistance states, reflecting the greater sensitivity of deeply scaled devices to fabrication-induced fluctuations. Nevertheless, all device dimensions retain clear hysteretic switching and self-rectifying behavior, demonstrating the robustness of the AlScN FeD architecture under aggressive lateral scaling.

To enable direct comparison across device dimensions, all measurements were performed on devices integrated within the same 8-B crossbar architecture, while varying the junction diameter from 1 μm to 100 nm. As the device diameter decreases from 1 μm to 100 nm, the absolute current scales proportionally with junction area, resulting in a substantial reduction in both ON- and OFF-state currents. After normalization by device area, the ON-state current densities from all device dimensions converge to a similar range at higher bias, indicating that the switching mechanism is governed by an area-dependent transport process. This behavior is consistent with polarization-controlled modulation of the tunneling barrier across the entire junction area and contrasts with filamentary resistive switching mechanisms, which typically exhibit a much weaker dependence on device area.

At lower bias, the OFF-state current densities do not fully overlap after area normalization. However, this apparent discrepancy primarily originates from the measurement floor of the source-measure unit, indicated by the dashed circle in Fig. 3h. Because the measured leakage currents approach the instrument resolution limit for all device dimensions, area normalization artificially broadens the OFF-state current density distribution of the smaller devices. Consequently, the true OFF-state current density of the scaled devices is expected to be lower than measured, suggesting that the intrinsic ON/OFF ratio may be even larger than the experimentally extracted values.

The statistical distributions of ON/OFF ratio and rectification ratio are summarized in Fig. 3i and j for devices with diameters ranging from 100 nm to 1 μm. Both figures of merit decrease monotonically with diameter and their interquartile spread widens. At 100 nm the 28 % Sc devices have a median ON/OFF ratio close to unity, so a usable memory window at this dimension is obtained only for the 32 % Sc composition (median ≈ 10). The 100 nm data therefore establish that polarization-controlled switching survives to this dimension, but the rectification available there (median ≈ 10) is well below what selector-free reading of a large passive array requires, and the array demonstrations that follow accordingly use 200 nm cells. We return to this constraint quantitatively in the discussion of Fig. 4.

Figure 3k presents the cumulative probability distributions of ON-state and OFF-state current densities for the three device dimensions. Distinct separation between the ON and OFF distributions is maintained across all sizes, indicating robust state distinguishability throughout the scaled device population. Although the statistical margin decreases somewhat with device scaling, the preserved distribution separation demonstrates that nanoscale AlScN FeDs remain suitable for high-density crossbar memory architectures.

Finally, pulse-mode measurements were performed to assess the frequency-dependent switching characteristics of the standalone AlScN FeD devices. Fig. 3l summarizes the extracted ON/OFF ratio as a function of pulse amplitude and frequency. As the programming frequency increases, a higher pulse voltage is required to maintain a large memory window, consistent with the finite switching dynamics of ferroelectric polarization reversal. The largest window, ON/OFF = $4.2 \times 10^3$, is obtained at 17.5 V and 1 kHz, and ON/OFF ratios above $10^2$ are maintained across the 12-19 V range at 1-10 kHz. At 100 kHz, however, the window falls to 13 even at 19 V and to below unity at 15 V, so the practical open-loop programming bandwidth of these cells is presently 10-50 kHz rather than the ~20 MHz implied by the 50 ns capacitor switching of Fig. 2h,i. We attribute this gap to the voltage division across the 4 nm $HfO_x$ interlayer, which reduces the field across the ferroelectric relative to the capacitor measurements, and to the finite RC time constant of the crossbar lines and probe parasitics; resolving it is a prerequisite for using these arrays at DRAM-like write speeds.

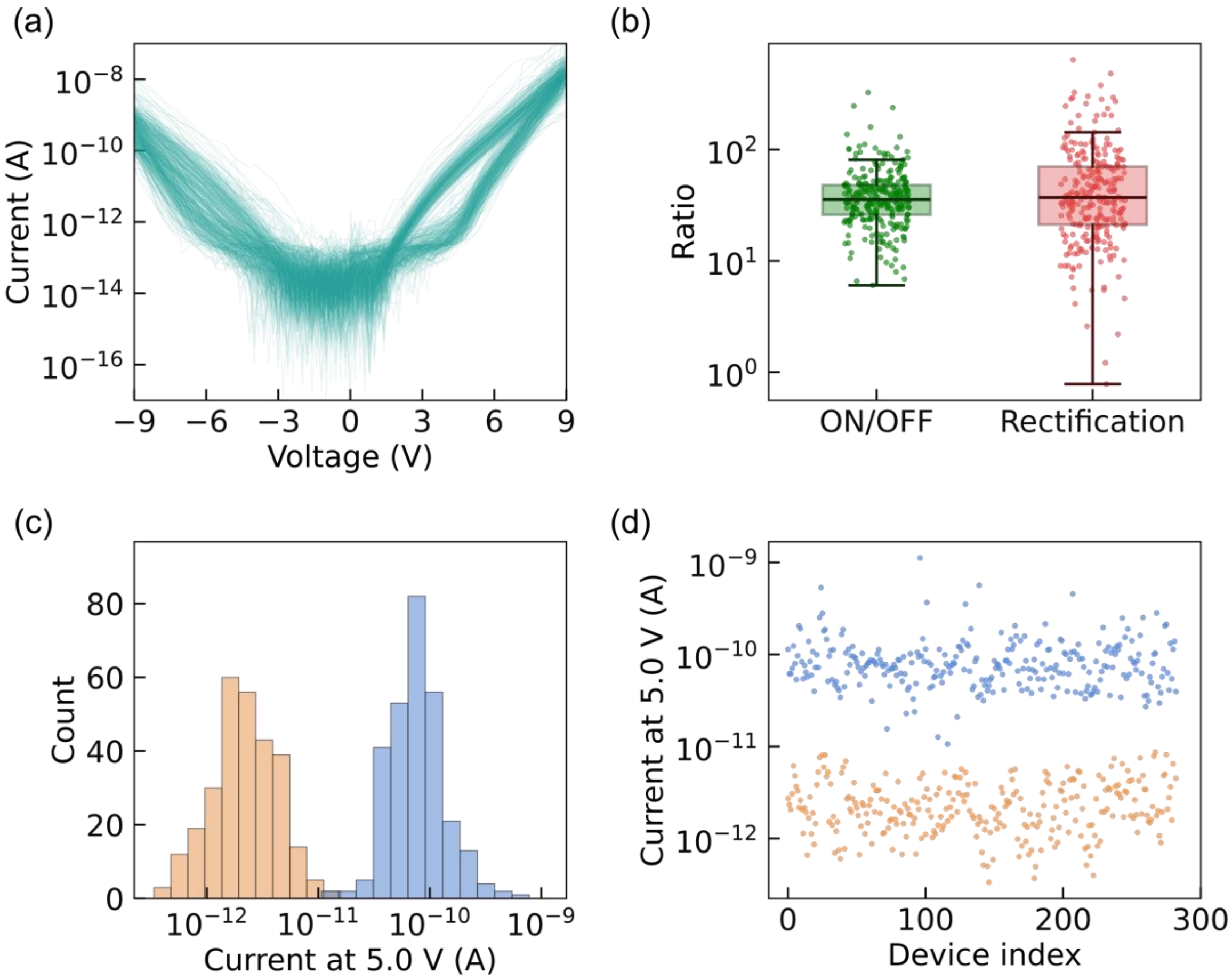

**Figure 4. Array-level uniformity of the 2-KB AlScN ferroelectric-diode crossbar array.** (a) Current–voltage characteristics of 283 cells randomly selected across the array, showing reproducible hysteretic switching and self-rectification. (b) Statistical distributions of the ON/OFF and rectification ratios extracted from the measured cells. (c) Histograms of the ON- and OFF-state currents measured at 5 V, showing clear separation between the two resistance-state distributions. (d) ON- and OFF-state currents plotted as functions of device index for the 283 measured cells, demonstrating uniform readout characteristics across the array.

### Array-Level Uniformity of a 2-KB AlScN Ferroelectric Diode Crossbar Array

Having established robust FeD operation across multiple device dimensions, we next evaluate the statistical uniformity of the full 2-KB crossbar array. A total of 283 devices - 1.7 % of the 16,384 cells - were randomly selected throughout the 128 × 128 array and electrically characterized to assess device-to-device variation at the array level. Figures 4a-d summarize the resulting switching and read-current statistics.

Figure 4a shows the I-V characteristics of the 283 randomly selected devices from the 2-KB crossbar array. All devices exhibit reproducible self-rectifying switching behavior with clearly distinguishable high- and low-resistance states. The statistical distributions of ON/OFF ratio and rectification ratio are summarized in Fig. 4b, yielding mean values of 42.8 and 60.8 with standard deviations of 33.2 and 74.2, and medians of 35.6 and 37.1. Both distributions are right-skewed and approximately log-normal: the coefficient of variation is 0.78 for the ON/OFF ratio and 1.22 for the rectification ratio, and a small fraction of cells has a rectification ratio below unity. Because sneak conductances add in parallel as a sum of 1/RR, the quantity that governs array read margin is the harmonic rather than the arithmetic mean of the rectification ratio, which for this distribution is approximately 22.

To further evaluate read-state distinguishability, Fig. 4c presents histograms of the ON- and OFF-state currents measured at 5 V. Two clearly separated distributions are observed with minimal overlap, confirming reliable discrimination between memory states during array readout. The same data are plotted as a function of device index in Fig. 4d, where both ON- and OFF-state currents remain remarkably stable across the 283 randomly selected cells. The limited spread of the read currents highlights the excellent array-level uniformity of the AlScN FeD platform and supports its suitability for large-scale selector-free memory integration.

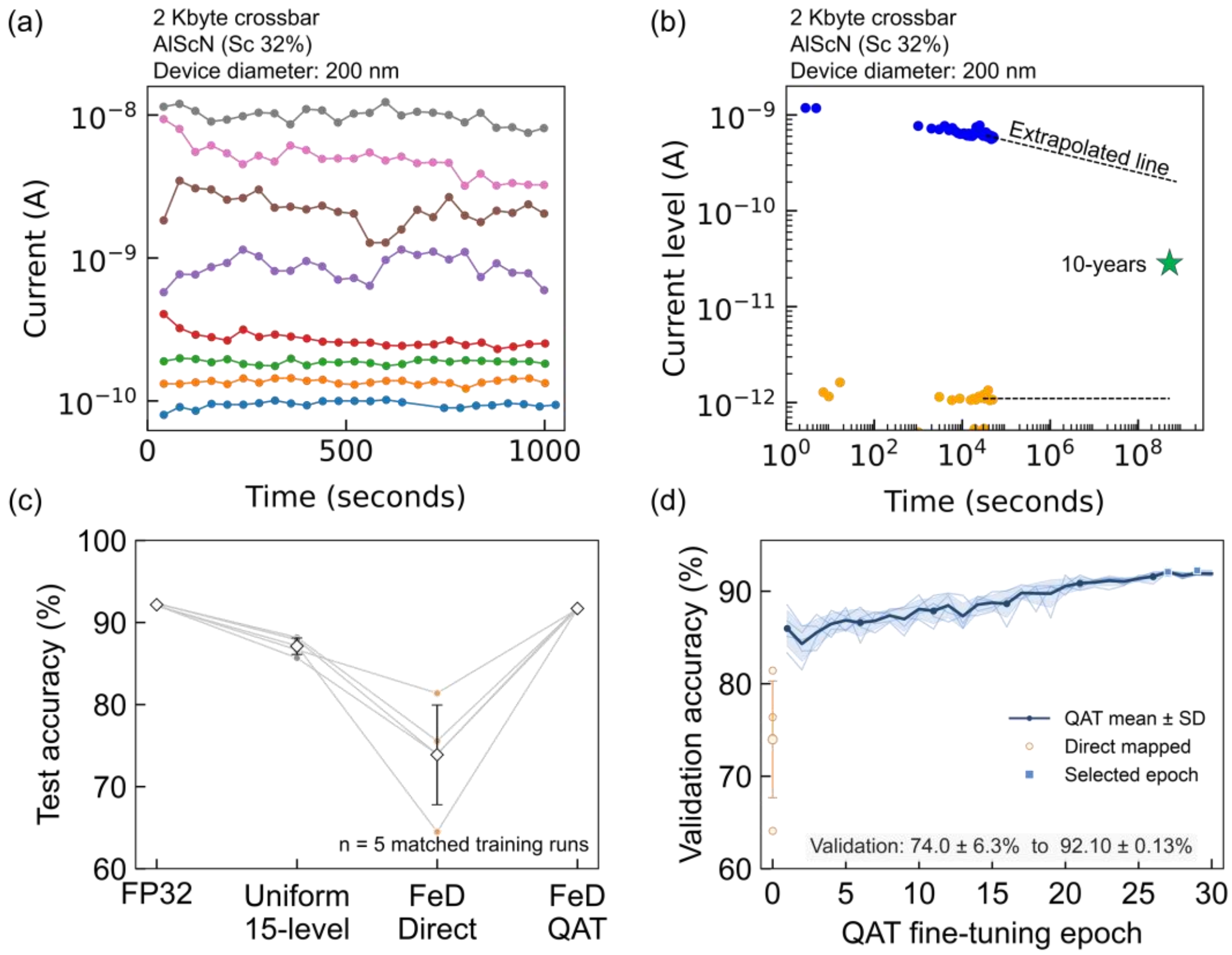

**Figure 5. Multilevel retention and hardware-aware neural-network inference using AlScN ferroelectric diodes.** (a) Retention of eight programmed current levels in a representative 200 nm device from the 2-KB crossbar array, measured for 1000 s. (b) Extended retention of the highest- and lowest-conductance states for more than $5 \times 10^4$ s. The dashed lines show linear extension, and the star denotes the projected current level after ten years. (c) CIFAR-10 test accuracy of a ResNet-20 model using full-precision weights, uniformly spaced signed 15-level weights, directly mapped experimental FeD levels, and experimental FeD levels with measured-level quantization-aware training (QAT). Symbols and error bars represent the mean and standard deviation from five matched training runs. (d) Validation accuracy as a function of QAT fine-tuning epoch. Open symbols indicate direct-mapping results, the line and shaded region show the QAT mean and standard deviation, and the filled square identifies the selected model epoch.

## Multilevel Retention and Inference Emulation in AlScN Ferroelectric Diode Arrays

To evaluate the stability of multilevel memory operation, retention measurements were performed on the programmed resistance states within the 2-KB crossbar array. Figure 5a shows the temporal evolution of eight distinct current levels measured from a representative 200 nm device. The eight levels span roughly two decades of read current, from ~$1 \times 10^{-10}$ to ~$1 \times 10^{-8}$ A, corresponding to a mean level spacing of a factor of ~1.9, and all eight remain resolvable over $10^3$ s. We note that this spacing is smaller than the cell-to-cell spread of the ON-state read current in Fig. 4d, so three-bit operation across the array would require write-verify programming rather than the open-loop pulse programming used here.

To further assess long-term retention, the highest- and lowest-conductance states were monitored for more than $5 \times 10^4$ s, as shown in Fig. 5b. Both states exhibit only modest current drift throughout the measurement period. A power-law extrapolation of the two extreme states over the measured window projects a memory window of order $10^2$ at $3.15 \times 10^8$ s (ten years) at room temperature. Because this extrapolation spans nearly four decades beyond the measured data and was acquired at room temperature only, it should be treated as an upper bound; accelerated retention at 85 °C and 125 °C, and eight-level retention over the same window, remain to be established. Within these limits, the data show that aggressive lateral scaling does not compromise the retention of the two extreme states in AlScN FeD arrays.

To evaluate the potential impact of the experimentally measured multilevel states on neural-network inference, we performed a hardware-aware simulation using the retained current levels from the AlScN FeD array. Eight measured 6 V retention-current states were converted into a signed 15-level weight codebook using a differential-pair mapping, preserving the experimentally measured, non-ideal level spacing rather than assuming uniformly spaced conductance states (Fig. 5c). When directly mapped to a CIFAR-10 ResNet-20 model, the measured FeD levels caused a substantial accuracy reduction due to nonuniform state spacing; however, measured-level quantization-aware training (QAT) recovered the test accuracy to 91.70 ± 0.12 %, within 0.46 percentage points of the FP32 baseline of 92.16 ± 0.14 % (Fig. 5d). These results suggest that although direct use of the raw device levels is insufficient for high-accuracy inference, algorithm-device co-optimization through QAT can compensate for nonideal multilevel conductance distributions in AlScN FeD memories. We stress the scope of this simulation: it propagates the measured level positions of a single cell, and it does not include the cell-to-cell variability of Fig. 4b, cycle-to-cycle programming noise, read disturb, retention drift over the inference interval, or array sneak currents. The accuracy reported here is therefore an upper bound on what the present array would deliver in hardware, and the gap between the two is set by precisely the non-idealities quantified earlier in this work.

Finally, Table 1 compares the present work with previously reported ferroelectric and emerging nonvolatile memory technologies. While several platforms have demonstrated either aggressive scaling, large array size, or selector-free operation individually, simultaneously achieving nanoscale dimensions, high ON/OFF ratio, selector-free functionality, and CMOS BEOL compatibility remains challenging. The AlScN FeD platform presented here combines a compact two-terminal architecture with intrinsic self-rectification, enabling a 2-KB passive crossbar array without additional selector devices. Furthermore, the 200 nm cells used in the array correspond to a junction area of 0.0314 $\mu m^2$ (0.00785 $\mu m^2$ for the 100 nm devices characterized in the 8-B arrays), which to our knowledge is the smallest cell area yet reported for a selector-free ferroelectric-diode crossbar. The array capacity demonstrated here matches that of our previously reported 5 μm-cell AlScN FeD crossbar; the advance is therefore a ~625× reduction in cell area at equal capacity, together with three-bit operation and nanosecond polarization reversal. Taken with the peak pulsed ON/OFF ratio of $4.2 \times 10^3$ and the demonstrated multilevel operation, these results define the current state of the art for scaled, selector-free ferroelectric-diode memory, while the rectification ratio measured at the array level sets the near-term limit on how far this architecture can be extended without a selector.

| Material | Structure | Non-volatile | Selector required? | Cell Size (µm²) | ON/OFF | Rectification | Multistate | BEOL-compatible | Array Size | Reference |
|---|---|---|---|---|---|---|---|---|---|---|
| HZO | 1T1C | Yes | Yes | 0.0023 [a] | N/A | N/A | ~ | Yes | 4 GB | IEDM 2023 [13] |
| HZO / IGZO | 1 FeD | Yes | No | 0.25 | $3 \times 10^5$ | $10^0$ * | ~ | No | ~ [b] | VLSI 2021 [14] |
| HZO | 1 FeD | Yes | No | 0.0249 [c] | $1 \times 10^4$ | $10^0$ * | ~ | Yes | 256 B | Nature Comm. 2020 [15] |
| HZO | 1T1C | Yes | Yes | ~ [d] | N/A | N/A | ~ | Yes | 8 KB | VLSI 2020 [16] |
| HZO | 1 FeD | Yes | No | ~ [e] | $1 \times 10^2$ | $10^1$ * | ~ | ~ | 1 KB | IEDM 2023 [17] |
| IGZO / $TeO_x$ | 2T0C | No | Yes | ~ [f] | N/A | N/A | 4 | Yes | 4 B [g] | IEDM 2024 [18] |
| C:$Ge_2Sb_2Te_5$ | 1T1R | Yes | Yes | ~ | $1.3 \times 10^1$ | N/A | 4 | ~ | 36 KB [h] | IEDM 2024 [19] |
| HZO | 1 FeD | Yes | No | ~ | $1 \times 10^3$ | $10^2$-$10^3$ * | ~ | No [i] | ~ [j] | IEEE EDL 2026 [20] |
| Al:HZO | 1T1C | Yes | Yes | 0.01 – 0.04 | N/A | N/A | ~ | Yes | 128 B | IEDM 2025 [21] |
| **AlScN** | **1 FeD** | **Yes** | **No** | **0.00785 [k]** | **$4.2 \times 10^3$ [l]** | **1481 (RR) [m]** | **8 [n]** | **Yes** | **2 KB** | **This Work** |

NL = nonlinearity, I(Vr)/I(Vr/2); RR = rectification ratio, I(+V)/I(−V). N/A = not applicable to the cell type; ~ = not reported.
*Estimated by the present authors from published I-V curves (order-of-magnitude only)
[a] $4F^2$ at 48 nm pitch (F = 24 nm)
[b] Array dimensions not stated; 250 cells measured.
[c] 19 nm TiN thickness × 1.31 µm hole perimeter.
[d] Capacitor size
[e] Reported only as 0.26 $F^2$.
[f] Cell area not reported ; transistor channels are W/L = 45/3 µm (read) and 15/3 µm (write)
[g] 4 × 4 array at 2 bit/cell
[h] 288 Kb capacity (2 × 144 Kb cores).
[i] HZO crystallized by RTA at 600 °C.
[j] No array demonstrated; five-layer device stack only.
[k] d = 100 nm, as used in the 2-kB array (0.00785 µm²)
[l] Pulsed write, 1 kHz / 17.5 V, read at 6 V. DC array-level ON/OFF: 275 (8-B, 28% Sc), 156 (8-B, 32% Sc), 48 (2-kB).
[m] DC, 8-B array, 32% Sc; 2-kB array average 232.
[n] 3-bit; all 8 states stable > $10^3$ s in the 2-kB array.

**Table 1. Benchmarking table.** Comparison of the FE-diode in this work to other FE-diodes and ferroelectric tunneling junctions (FTJs) based on various ferroelectric materials.[13-21] Key metrics include CMOS BEOL compatibility, selector requirement, device footprint, ON/OFF ratio, rectification ratio (RR), switching speed, and array-level integration. FE-diodes provide intrinsic self-rectifying characteristics, enabling selector-free memory arrays, whereas FTJs may require additional selector devices to suppress sneak-path currents.

## CONCLUSION

We have demonstrated selector-free AlScN ferroelectric-diode memories spanning standalone devices, 8-byte arrays and a 2-kilobyte (128 × 128) passive crossbar built on 200 nm cells, with standalone and 8-byte-array devices scaled to 100 nm. The 10 nm $Al_{0.64}Sc_{0.36}N$ layer retains ferroelectric switching and reverses its polarization under voltage pulses of 50 ns full width at half maximum, at a pulse-energy density of ~16 pJ $\mu m^{-2}$ corresponding to ~0.5 pJ per 200 nm cell.

After optimizing the Sc content to 32 %, junctions of 200 nm diameter retain nonvolatile resistance switching and intrinsic self-rectification, with a peak pulsed ON/OFF ratio of $4.2 \times 10^3$ at 17.5 V and 1 kHz and array-level DC values of 43 ± 33 (ON/OFF) and 61 ± 74 (rectification) at 5 V across 283 randomly addressed cells. Eight conductance levels remain distinguishable for more than $10^3$ s in a 200 nm array cell, and the two extreme states are stable beyond $5 \times 10^4$ s. Incorporating the measured, nonuniformly spaced levels into a ResNet-20 model, quantization-aware training recovers a CIFAR-10 accuracy of 91.7 %, within 0.5 percentage points of the full-precision baseline. At a junction area of 0.0314 $\mu m^2$ the cell is ~625× smaller than in previously reported kilobyte-scale AlScN FeD crossbars of equal capacity.

Three limits define the work that follows. First, the rectification ratio available at 200 nm - a harmonic mean of ~22 across the array - supports selector-free reading of sub-arrays of order 26 × 26 at a 10 % worst-case read margin, so extending random access across the full 2-kB array, let alone to megabit capacities, requires the rectification ratio to be raised by roughly two orders of magnitude. Interlayer engineering, lower-bias reading in the direct-tunnelling-limited regime, and thicker ferroelectrics at fixed cell area are the obvious levers. Second, the practical open-loop programming rate of these cells, 10-50 kHz, is more than three orders of magnitude slower than the 50 ns polarization reversal measured on capacitors of the same film; closing that gap is a question of series electrostatics and interconnect parasitics rather than of intrinsic ferroelectric dynamics, and it is tractable. Third, write endurance and read/write disturb at these dimensions remain to be established, as does retention of all eight levels at elevated temperature. None of these limits is peculiar to the wurtzite system, and the combination demonstrated here - three-bit storage, nanosecond intrinsic switching, sub-0.05 $\mu m^2$ cells and a thermal budget compatible with CMOS back-end-of-line integration - makes AlScN ferroelectric diodes a credible basis for dense, selector-free nonvolatile memory and for in-memory computing hardware.

## METHODS

**Al/AlScN/Al deposition.**

$Al/Al_{1-x}Sc_xN/Al$ stacks were deposited on 100 mm c-axis-oriented sapphire substrates with a 0.2° ±0.1° miscut toward the M-plane. The multilayer stacks were grown using an Evatec CLUSTERLINE® 200 II physical vapor deposition system with a target-to-substrate distance of 88.5 mm. All layers were deposited sequentially without breaking vacuum to minimize oxidation and contamination at the Al/AlScN interfaces. The 50 nm-thick bottom and top Al electrodes were deposited at 150 °C using an Al target power density of 12.7 $W/cm^2$. An Ar flow rate of 20 sccm produced a process pressure of approximately $1.1\times10^{-3}$ mbar. Under these conditions, the Al deposition rate was approximately 1.3 nm/s, and the films exhibited a preferred <111> orientation.

The $Al_{1-x}Sc_xN$ layers were deposited on the bottom Al electrode by reactive co-sputtering from separate Al and Sc targets at 350 °C. Three compositions, $Al_{0.72}Sc_{0.28}N$, $Al_{0.68}Sc_{0.32}N$, and $Al_{0.64}Sc_{0.36}N$, were obtained by adjusting the Sc cathode power between 450 and 555 W while maintaining the Al cathode power at 900 W. For the AlScN films, the $N_2$ flow rate was maintained at 30 sccm, resulting in a process pressure of approximately $8.0\times10^{-4}$ to $8.5\times10^{-4}$ mbar[22]. The film thickness was controlled by varying the deposition duration, with a nominal AlScN deposition rate of approximately 0.25 nm/s.

**Crossbar Device fabrication.**

The fabrication process is adapted from previous works[23, 24]. The crossbar arrays were fabricated from an in-situ-deposited Al/AlScN/Al stack on sapphire. Au alignment markers were firstly deposited and patterned to enable accurate overlay during the following electron-beam lithography steps. The bottom word-line structures were defined by electron-beam lithography using a $SiO_2$ hard mask. The patterned $SiO_2$ hard mask and the underlying Al/AlScN/Al layers were sequentially etched by RIE and IBE, respectively, to isolate the bottom Al electrodes and define adjacent word lines. The remaining $SiO_2$ hard mask was then removed by an additional RIE process. A ICP-RIE process with mixture of $BCl_3/Cl_2$ gas was used to remove the top Al cover. A high-quality $HfO_x$ interlayer was deposited by atomic layer deposition (ALD) at 150 °C. A 200 nm-thick $SiO_2$ passivation layer was subsequently deposited by plasma-enhanced chemical vapor deposition to electrically isolate the devices and planarize the array surface. Contact vias were defined by electron-beam lithography, followed by RIE of the $SiO_2$ passivation layer to expose the underlying device electrodes. Finally, 30 nm Ti/200 nm Au upper metal interconnects and contact pads were deposited and patterned to form bit lines oriented orthogonally to the bottom word lines, completing the crossbar-array architecture.

**Electrical Characterization.**

Electrical measurements were performed in ambient air at room temperature using a Lake Shore probe station and a Keithley 4200A semiconductor parameter analyzer. Voltage was applied to the Al bottom electrode while the Ti/Au top electrode was grounded. The ON/OFF and rectification ratios were extracted at 5 V. Ferroelectric J-E measurements were performed at 10 kHz on devices with a radius of 5 μm. PUND and pulse-switching measurements were conducted using a pulse measurement unit, with the switching charge obtained by integrating the P-U and N-D current differences. Switching was evaluated using pulses with full widths at half maximum down to 50 ns. The switching-energy density was calculated by integrating $V(t)*I_{sw}(t)$ and normalizing by the device area. Array-level statistics were obtained from 283 randomly selected cells of the 128 × 128 (2 kB) array. During array measurements the selected word line was biased and the selected bit line grounded, with all unselected word and bit lines left floating; no external current compliance was applied. Multilevel retention was measured at a read bias of 6 V, with the eight states written by single voltage pulses of increasing amplitude.

The device architecture and interface connectivity were characterized by cross-sectional scanning transmission electron microscopy (STEM) and energy-dispersive X-ray spectroscopy (EDS). STEM imaging and EDS mapping were performed on an aberration-corrected JEOL NEOARM operated at 200 kV and equipped with two large-solid-angle SDDs. A probe convergence semi-angle of 25 mrad and a camera length of 4 cm were used. Cross-sectional TEM lamellae were prepared by focused ion beam (FIB) milling using a Ga+ ion Strata DB235, with final thinning at 5 kV to minimize surface damage.

**Simulation.**

Hardware aware neural network simulations were performed using experimentally measured retention currents from 8 programmed states of a 32% Sc AlScN FeD in the 2-KB crossbar array. Currents were measured at 6 V and interpolated in log time and log current space over the common retention window. Signed weights were represented using differential FeD pairs, with the lowest current state serving as the reference, producing 15 signed levels indexed from -7 to +7 while preserving the experimentally measured nonuniform level spacing.

Inference was evaluated using CIFAR 10 and a ResNet 20 model. A fixed split of 45,000 training and 5,000 validation images was used, with a 10,000 image test set held out for final evaluation. Five matched random seeds were used for all conditions. FP32 models were trained for 200 epochs using stochastic gradient descent and cosine learning rate decay. Four conditions were compared: FP32, uniformly spaced 15 level quantization, direct measured FeD mapping, and measured FeD quantization aware training. QAT was initialized from the selected FP32 checkpoints and fine tuned for 30 epochs using fixed layer scale factors and a straight through estimator. Reported accuracies are the mean and sample standard deviation across five seeds.

## AUTHOR INFORMATION

### Author Contributions

Kwan-Ho Kim, Zekun Hu, and Krishna Karthikeya Chemudupati contributed equally to this work.

D.J., R.O., K.-H.K., Z.Hu and K.C. conceived the project. K.-H.K., Z.Hu and K.C. designed the experiments and fabricated the standalone ferroelectric diodes, 8-byte arrays, and 2-kilobyte crossbar arrays under the supervision of D.J.. Y.Z. and Z.Hu deposited the AlScN film. K.-H.K. K.C., T.J. and Z.Hu performed the electrical measurements, including J–E, C–V, PUND, endurance, nanosecond switching, DC and pulse-mode I–V characterization, device-scaling measurements, array-level statistical analysis, multilevel programming, and retention measurements. K.-H.K. K.C., T.J., Z.Hu, Z.Han, and D.K.P. analyzed the electrical data and switching characteristics. R.K.R. performed the cross-sectional TEM and EDS characterization under the supervision of E.S.. K.C. and T.J. developed and performed the neural-network inference simulations and measured-level quantization-aware training. D.K.P. contributed to electrical-data analysis and interpretation. K.-H.K. and Z.Hu prepared the initial manuscript, with K.C. and T.J. contributing to the hardware-aware inference section. D.J., R.O., and E.S. supervised the research.

All authors discussed the results, reviewed the manuscript, and approved its final version.

## ACKNOWLEDGEMENT

We thank Haiwen Zhang at University of Pennsylvania for initial assistance with data collection. We also acknowledge the staff of the Quattrone Nanofabrication Facility at the University of Pennsylvania for their support and helpful suggestions.

D.J.acknowledges primary support from the Office of Naval Research (ONR) Nanoscale Computing and Devices program (No. N00014–24–1–2131). D.J. also acknowledge support from the Air Force Office of Scientific Research (AFOSR) GHz-THz Program (Grant No. FA9550–23–1–0391) and from the National Science Foundation, Future of Semiconductors (FuSe) program: award number 2328743. A portion of the sample fabrication, assembly, and characterization were carried out at the Singh Center for Nanotechnology at the University of Pennsylvania, which is supported by the National Science Foundation (NSF) National Nanotechnology Coordinated Infrastructure Program grant NNCI-1542153. Additional support for the Nanoscale Characterization Facility at the Singh Center was provide by the NSF through the University of Pennsylvania Materials Research Science and Engineering Center (MRSEC) (DMR-2309043).

## Competing-interest declaration

D.J., R.H.O. and E.A.S. are co-founders of Agni Semiconductor, which is commercializing AlScN ferroelectric memory, and this is directly relevant to the subject matter of the paper

## REFERENCES

(1) Pal, S.; Mallik, A.; Gupta, P. System technology co-optimization for advanced integration. *Nature Reviews Electrical Engineering* **2024**, *1* (9), 569-580. DOI: 10.1038/s44287-024-00078-x.

(2) Lu, A.; Lee, J.; Kim, T.-H.; Karim, M. A. U.; Park, R. S.; Simka, H.; Yu, S. High-speed emerging memories for AI hardware accelerators. *Nature Reviews Electrical Engineering* **2024**, *1* (1), 24-34. DOI: 10.1038/s44287-023-00002-9.

(3) Ambrogio, S.; Narayanan, P.; Okazaki, A.; Fasoli, A.; Mackin, C.; Hosokawa, K.; Nomura, A.; Yasuda, T.; Chen, A.; Friz, A.; et al. An analog-AI chip for energy-efficient speech recognition and transcription. *Nature* **2023**, *620* (7975), 768-775. DOI: 10.1038/s41586-023-06337-5 From NLM PubMed-not-MEDLINE.

(4) Ielmini, D.; Wong, H. S. P. In-memory computing with resistive switching devices. *Nature Electronics* **2018**, *1* (6), 333-343. DOI: 10.1038/s41928-018-0092-2.

(5) Wan, W.; Kubendran, R.; Schaefer, C.; Eryilmaz, S. B.; Zhang, W.; Wu, D.; Deiss, S.; Raina, P.; Qian, H.; Gao, B.; et al. A compute-in-memory chip based on resistive random-access memory. *Nature* **2022**, *608* (7923), 504-512. DOI: 10.1038/s41586-022-04992-8 From NLM PubMed-not-MEDLINE.
(6) Alexander Flocke, a. T. G. N. Fundamental Analysis of Resistive Nano-Crossbars for the Use in Hybrid Nano/CMOS-Memory. *IEEE ESSCIRC* **2007**.
(7) Liang, J.; Wong, H. S. P. Cross-Point Memory Array Without Cell Selectors—Device Characteristics and Data Storage Pattern Dependencies. *IEEE Transactions on Electron Devices* **2010**, *57* (10), 2531-2538. DOI: 10.1109/ted.2010.2062187.
(8) Jiantao, Z.; Kuk-Hwan, K.; Wei, L. Crossbar RRAM Arrays: Selector Device Requirements During Read Operation. *IEEE Transactions on Electron Devices* **2014**, *61* (5), 1369-1376. DOI: 10.1109/ted.2014.2310200.
(9) Feng, G.; Zhu, Q.; Liu, X.; Chen, L.; Zhao, X.; Liu, J.; Xiong, S.; Shan, K.; Yang, Z.; Bao, Q.; et al. A ferroelectric fin diode for robust non-volatile memory. *Nat Commun* **2024**, *15* (1), 513. DOI: 10.1038/s41467-024-44759-5 From NLM PubMed-not-MEDLINE.
(10) Kim, K. H.; Karpov, I.; Olsson, R. H., 3rd; Jariwala, D. Wurtzite and fluorite ferroelectric materials for electronic memory. *Nat Nanotechnol* **2023**, *18* (5), 422-441. DOI: 10.1038/s41565-023-01361-y From NLM PubMed-not-MEDLINE.
(11) Kim, K. H.; Oh, S.; Fiagbenu, M. M. A.; Zheng, J.; Musavigharavi, P.; Kumar, P.; Trainor, N.; Aljarb, A.; Wan, Y.; Kim, H. M.; et al. Scalable CMOS back-end-of-line-compatible AlScN/two-dimensional channel ferroelectric field-effect transistors. *Nat Nanotechnol* **2023**, *18* (9), 1044-1050. DOI: 10.1038/s41565-023-01399-y From NLM PubMed-not-MEDLINE.
(12) Kim, K. H.; Han, Z.; Zhang, Y.; Musavigharavi, P.; Zheng, J.; Pradhan, D. K.; Stach, E. A.; Olsson, R. H., 3rd; Jariwala, D. Multistate, Ultrathin, Back-End-of-Line-Compatible AlScN Ferroelectric Diodes. *ACS Nano* **2024**, *18* (24), 15925-15934. DOI: 10.1021/acsnano.4c03541 From NLM PubMed-not-MEDLINE.
(13) Ramaswamy, N.; Calderoni, A.; Zahurak, J.; Servalli, G.; Chavan, A.; Chhajed, S.; Balakrishnan, M.; Fischer, M.; Hollander, M.; Ettisserry, D. P.; et al. NVDRAM: A 32Gb Dual Layer 3D Stacked Non-volatile Ferroelectric Memory with Near-DRAM Performance for Demanding AI Workloads. In 2023 International Electron Devices Meeting (IEDM), 2023.
(14) al., H. B. e. Ferroelectric Diodes with Sub-ns and Sub-fJ Switching and Its Programmable Network for Logic-in-Memory Applications. *2021 VLSI* **2021**.
(15) Luo, Q.; Cheng, Y.; Yang, J.; Cao, R.; Ma, H.; Yang, Y.; Huang, R.; Wei, W.; Zheng, Y.; Gong, T.; et al. A highly CMOS compatible hafnia-based ferroelectric diode. *Nat Commun* **2020**, *11* (1), 1391. DOI: 10.1038/s41467-020-15159-2 From NLM PubMed-not-MEDLINE.
(16) al., J. O. e. SoC compatible 1T1C FeRAM memory array based on ferroelectric Hf0.5Zr0.5O2. *2020 VLSI* **2020**.

(17) G., T. e. a. First Demonstration of a Bayesian Machine based on Unified Memory and Random Source Achieved by 16-layer Stacking 3D Fe-Diode with High Noise Density and High Area Efficiency. In 2023 International Electron Devices Meeting (IEDM), 2023.
(18) Su, Y.; Liu, T.; Tang, J.; Li, Y.; An, R.; Du, Y.; Tang, Z.; Zhang, Y.; Fan, Y.; He, Y.; et al. Complementary Oxide Semiconductor-Based 2T0C DRAM Macro with CFET Peripherals Using TeOx-PFET/IGZO-NFET for 3D Memory Integration. In 2024 IEEE International Electron Devices Meeting (IEDM), 2024.
(19) Yan, L.; Li, Y.; Li, X.; Pan, Z.; Wang, Z.; Wang, X.; Wang, B.; Zhan, Z.; Tang, X.; Tao, Y.; et al. Neural Manifold Learning Based on 40 nm Dual-Mode PCM Compute-in-Memory Chip with Hardware Adaptive Drift Compensation. In 2024 IEEE International Electron Devices Meeting (IEDM), 2024.
(20) Zhou, W.; Yan, R.; Xue, R.; Chan, M.; Liu, X. All-ALD Sequential Monolithic Integration of Selector-Free Five-Layer Ferroelectric Diodes for 3-D Cross-Point Memory. *IEEE Electron Device Letters* **2026**, *47* (8), 1663-1666. DOI: 10.1109/led.2026.3704279.
(21) Müller, F.; Iurchuk, V.; Viegas, A.; Yadav, N.; Syndikus, J.; Seidel, R.; Lehninger, D.; Hoffmann, R.; Olivo, R.; Mutter, J.; et al. Highly Scaled FRAM Technology in 22FDX® Achieving Sub 1 V Operation. In 2025 IEEE International Electron Devices Meeting (IEDM), 2025.
(22) Zhang, Y.; Smith, W. J.; Esteves, G.; Leblanc, C.; Stach, E. A.; Beechem, T. E.; Olsson, R. H. Tuning Breakdown-to-Coercive Field Ratio in Ultra-Thin Al1–xScxN Films via Reactive Nitrogen Atmosphere. *IEEE Transactions on Electron Devices* **2026**, *73* (8), 4856-4864. DOI: 10.1109/ted.2026.3704526.
(23) Han, Z.; Chen, C.-C.; Pradhan, D. K.; Moore, D. C.; Gudavalli, R.; Yang, X.; Kim, K.-H.; Cho, H.; Anderson, Z.; Ware, S.; et al. Kilobyte-scale, selector-free, thermally robust AlScN ferroelectric diode crossbar arrays. *Device* **2026**, *4* (2). DOI: 10.1016/j.device.2025.100974.
(24) Hu, Z.; Cho, H.; Rai, R. K.; Bao, K.; Zhang, Y.; Qu, Z.; He, Y.; Ji, Y.; Leblanc, C.; Kim, K. H.; et al. Demonstration of Highly Scaled AlScN Ferroelectric Diode Memory with a Storage Density of >100 Mbit/mm(2). *Nano Lett* **2025**, *25* (37), 13748-13755. DOI: 10.1021/acs.nanolett.5c02961 From NLM PubMed-not-MEDLINE.

# Supporting information for:

# Multilevel AlScN Ferroelectric Diode Crossbar Arrays with Sub-200 nm Cells for Selector-Free Nonvolatile Memory

Kwan-Ho Kim,[1†] Zekun Hu,[1†] Krishna Karthikeya Chemudupati,[1,2†] Taarana Jammula,[1] Yinuo Zhang,[1] Rajeev Kumar Rai,[3] Dhiren K. Pradhan,[1] Zirun Han,[1] Eric A. Stach,[3] Roy H. Olsson III,[1] and Deep Jariwala[1,*]

*[1]Department of Electrical and Systems Engineering, University of Pennsylvania, Philadelphia, Pennsylvania 19104, United States*

*[2]Department of Physics and Astronomy, University of Pennsylvania, Philadelphia, Pennsylvania 19104, USA*

*[3]Department of Materials Science and Engineering, University of Pennsylvania, Philadelphia, Pennsylvania 19104, USA*

*[†]These authors equally contributed to this work.*

[*]Author to whom correspondence should be addressed: dmj@seas.upenn.edu

# Contents

## Note S1. Positive Voltage Transport Fitting.

This section evaluates the dominant transport behavior of the ferroelectric diodes in the positive voltage range. The low resistance state and high resistance state are described using Poole Frenkel emission and thermionic emission, respectively. The central result is that one shared parameter set describes the measured devices from the 2 kB array, supporting a reproducible change in transport behavior between the two resistance states.

### Poole Frenkel Transport in the Low Resistance State

In the LRS, polarization induced lowering of the injection barrier enables carriers to access trap states in AlScN. Transport through the ferroelectric is therefore treated as a bulk limited Poole Frenkel process. The Poole Frenkel current density is expressed as

$$J_{PF} \; = \; A_{PF}\, E \; exp[-\frac{q}{k_B T}(\Phi_t \; - \; \sqrt{\frac{qE}{\pi\varepsilon}})]$$

where $A_{PF}$ is an effective prefactor, E is the electric field, $\Phi_t$ is the trap activation energy, ε is the dielectric permittivity, q is the elementary charge, $k_B$ is the Boltzmann constant, and T is the temperature. The square root term represents the electric field induced lowering of the trap emission barrier. This model was applied to the positive voltage LRS return branch.

### Thermionic Emission in the High Resistance State

In the low bias HRS, the increased injection barrier suppresses carrier injection and the current is treated as interface limited thermionic emission. The thermionic emission current density is expressed as

$$J_{TI} \; = \; A^{*}T^{2} \; exp[-\frac{q}{k_B T}(\Phi_B \; - \; \sqrt{\frac{qE}{4\pi\varepsilon}})]$$

where $A^*$ is the effective Richardson constant and $\Phi_B$ is the effective carrier injection barrier. The square root term represents image force lowering of the interface barrier under the applied electric field. This model was applied to the positive voltage HRS from 0.25 to 4.5 V, before the rapid increase in current associated with polarization switching.

### Model Interpretation

For a film of thickness d, the average field is approximated as E = V/d. Both models are field activated because the applied field lowers an energy barrier, but the location and physical origin of that barrier are different. Poole Frenkel emission describes the field assisted release of a carrier from a charged trap within the AlScN bulk. Thermionic emission describes carrier injection over an interface barrier, with the field reducing that barrier through the image force.

$$\Delta\Phi_{PF} = \sqrt{\frac{qE}{\pi\varepsilon}}; \quad \Delta\Phi_{TI} = \sqrt{\frac{qE}{4\pi\varepsilon}} = \frac{1}{2}\Delta\Phi_{PF}$$

In the Poole Frenkel picture, the Coulombic potential that confines a carrier to a charged defect is tilted by the electric field. The barrier reduction is proportional to $\sqrt{E}$, so the emission probability and current increase exponentially with the square root of the field. After dividing out the explicit field factor, the model predicts that $ln\frac{J}{E}$ should vary approximately linearly with the square root of E over the field range in which one trap family and a nearly uniform field dominate. The slope is controlled by temperature and dielectric response, while the intercept combines the trap activation energy with the effective prefactor.

For thermionic emission, carriers must acquire sufficient thermal energy to cross the contact barrier. Image force lowering reduces the interface barrier by an amount proportional to $\sqrt{E}$. At fixed temperature, $ln\frac{J}{T^2}$ is therefore expected to vary approximately linearly with $\sqrt{E}$. For the same field and permittivity, the ideal thermionic barrier lowering is one half of the Poole Frenkel lowering because the thermionic expression contains $4\pi\varepsilon$ rather than $\pi\varepsilon$. This difference provides a useful physical constraint, although it is not by itself sufficient to identify a unique mechanism from a single current voltage sweep.

Reversal of the AlScN polarization changes the bound charge at the metal ferroelectric interfaces and consequently changes the local band bending and effective injection barrier. In the HRS orientation, the larger effective barrier suppresses low bias injection, making an interface limited thermionic description reasonable before switching begins. In the LRS orientation, the reduced injection barrier supplies carriers to the AlScN layer, allowing field assisted release from bulk traps to control the return branch and producing the stronger Poole Frenkel type field dependence.

The low bias thermionic fit is intentionally stopped before the pronounced current increase in the HRS sweep. Beyond this point, the device can no longer be treated as a stationary HRS with a single, bias independent interface barrier. Polarization reversal begins through domain nucleation and growth, so the bound interface charge, band bending, injection barrier, and internal field distribution can evolve during the voltage sweep. The measured current in this region may therefore contain a combination of steady conduction, capacitive displacement current, polarization switching current, and conduction through partially switched regions.

The fits establish distinct transport behavior in the selected low bias HRS and LRS return ranges. The higher voltage HRS current begins after polarization starts to evolve and therefore lies outside the stationary transport regimes described by these equations. Direct tunneling, trap assisted tunneling, field assisted trap emission, thermionic injection, and switching related current can have overlapping voltage dependences over a limited range. Separating these overlapping contributions remains a challenge, so the higher voltage HRS region is described here as a switching regime rather than assigned to a single steady state transport mechanism. Within the selected positive voltage intervals, the shared fits support an interface limited description for the low bias HRS and a bulk trap assisted description for the LRS return branch. The fitted quantities are therefore interpreted as effective transport parameters.

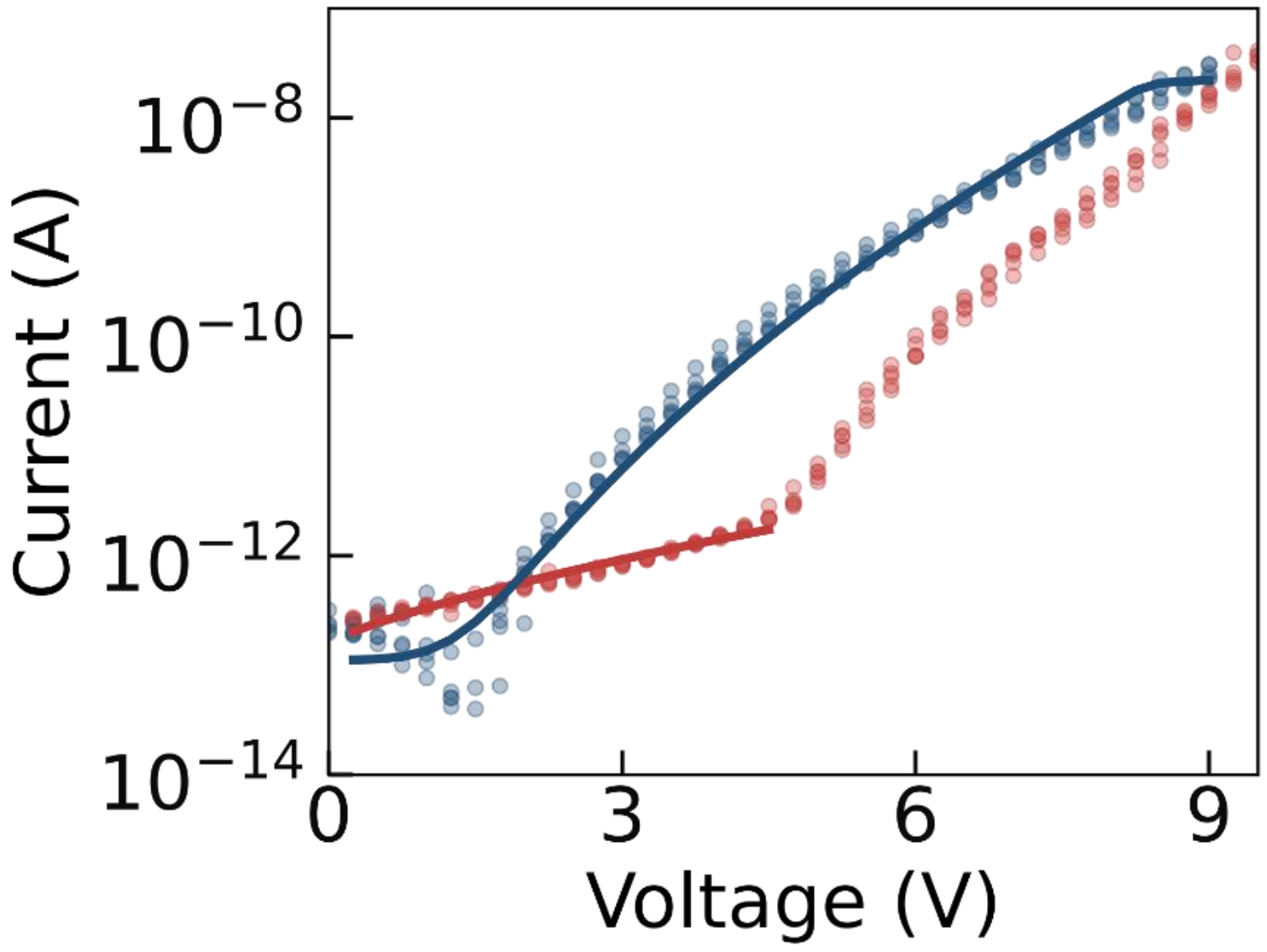

**Figure S1. Tunneling mechanism fitting.**

Positive voltage transport characteristics for representative devices from the 2 kB array. Deep blue and red markers show the measured LRS return and HRS branches, respectively. The deep blue solid line shows the shared Poole Frenkel fit to the LRS data. The red solid line shows the shared thermionic emission fit to the HRS data from 0.25 to 4.5 V. The complete positive going HRS data are displayed, while the thermionic fit is restricted to the low bias region before polarization switching.

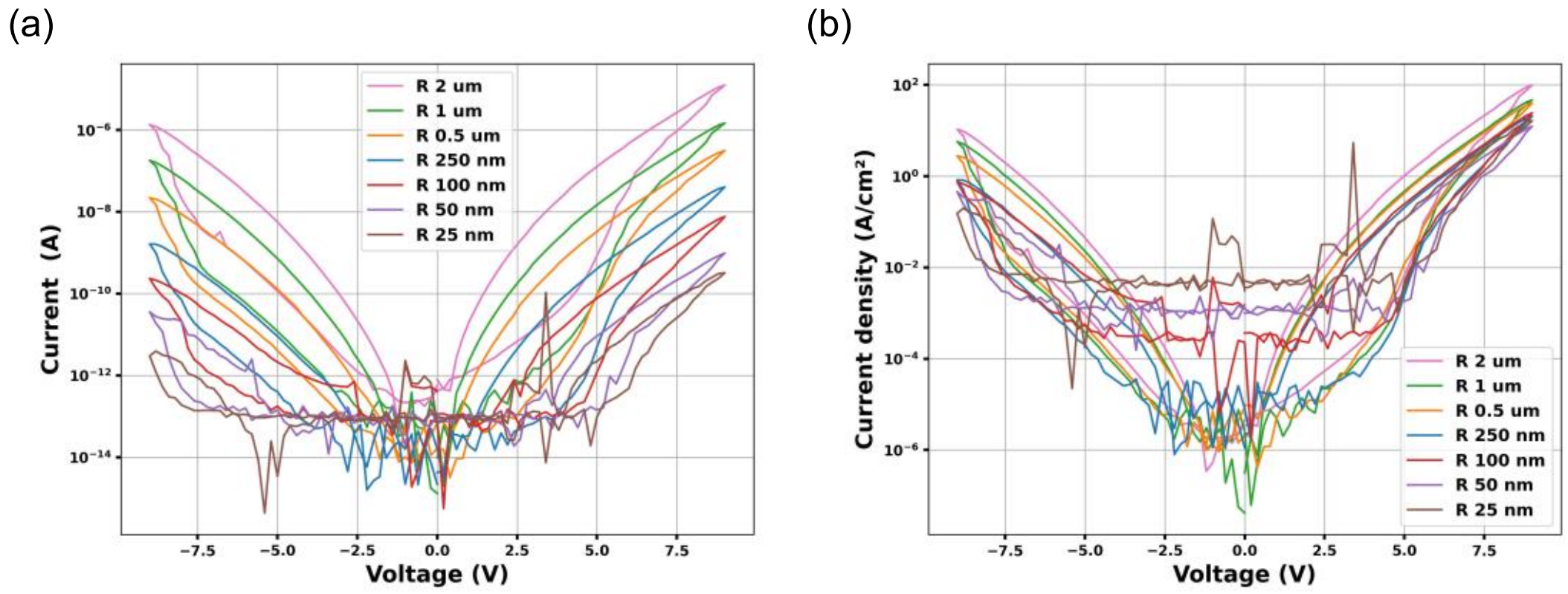

**Figure S2. Lateral scaling of standalone AlScN ferroelectric diodes.**

(a) Current–voltage and (b) area-normalized current density–voltage characteristics of devices comprising a 10 nm $Al_{0.68}Sc_{0.32}N$ ferroelectric layer and a 4 nm $HfO_2$ interlayer, with junction radii ranging from 2 μm to 25 nm. Hysteretic resistance switching and self-rectifying behavior persist upon lateral scaling, consistent with preserved polarization-controlled transport. The comparable current densities at high positive bias support predominantly area-dependent conduction, whereas the low-bias currents of the smallest devices approach the measurement floor.

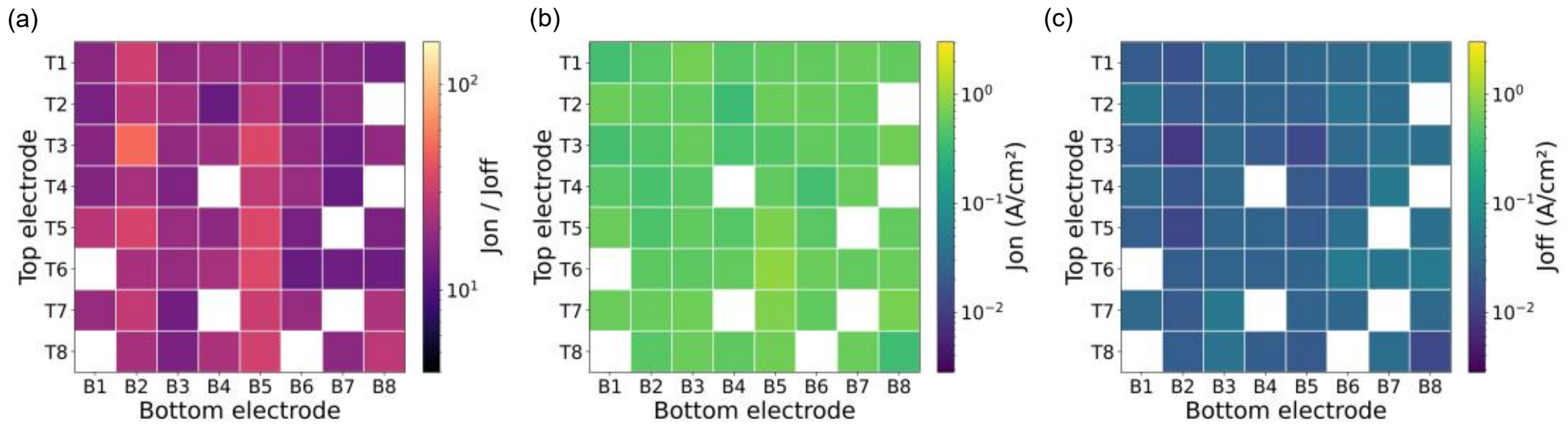

**Figure S3. Spatial maps of electrical characteristics in an 8 × 8 $Al_{0.64}Sc_{0.36}N$ ferroelectric-diode crossbar array.**

(a) ON/OFF current-density ratio ($J_{ON}/J_{OFF}$), (b) ON-state current density ($J_{ON}$), and (c) OFF-state current density ($J_{OFF}$). Rows and columns correspond to top electrodes T1–T8 and bottom electrodes B1–B8, respectively. Color scales are logarithmic. White cells indicate devices with currents outside the measurement range.